\documentclass[lettersize,journal]{IEEEtran}
\usepackage{amsmath,amsfonts}
\usepackage{xcolor}
\usepackage{algorithmic}
\usepackage{algorithm}
\usepackage{array}
\usepackage[caption=false,font=normalsize,labelfont=sf,textfont=sf]{subfig}
\usepackage{textcomp}
\usepackage{stfloats}
\usepackage{url}
\usepackage{verbatim}
\usepackage{graphicx}
\usepackage{booktabs}
\usepackage{caption}
\usepackage{longtable}
\usepackage{lipsum}
\usepackage{cite}
\begin{document}

\title{\fontsize{21}{25}\selectfont A new and flexible design method for Symmetric Quadrature Hybrid Couplers using Markov Chain Monte Carlo}

\author{Arjun Ghosh,~\IEEEmembership{Astronomy and Astrophysics, Raman Research Institute} \thanks{Arjun Ghosh, arjunghosh@ieee.org} \\
Ritoban Basu Thakur,~\IEEEmembership{Department of Physics, California Institute of Technology} \thanks{Ritoban Basu Thakur, ritoban@caltech.edu}}
        % <-this % stops a space
% \thanks{This paper was produced by the IEEE Publication Technology Group. They are in Piscataway, NJ.}% <-this % stops a space
% \thanks{Manuscript received April 19, 2021; revised August 16, 2021.}}

% The paper headers
% \markboth{IEEE TRANSACTIONS ON MICROWAVE THEORY AND TECHNIQUES,~Vol.~14, No.~8, November~2024}%
% {Shell \MakeLowercase{\textit{et al.}}: A Sample Article Using IEEEtran.cls for IEEE Journals}

% \IEEEpubid{0000--0000/00\$00.00~\copyright~2021 IEEE}
% Remember, if you use this you must call \IEEEpubidadjcol in the second
% column for its text to clear the IEEEpubid mark.

\maketitle

\begin{abstract}
Quadrature Hybrid Couplers (QHDC) are critical components in RF, mm-wave, and sub-mm wave astronomical instrumentation, where wideband performance with minimal passband ripple is essential. Traditional designs have been limited to 5-sections at most, due to computational limitations. In this work, we introduce a new analytical technique to design couplers with larger sections and improved performance. We do this by employing a Markov Chain Monte Carlo (MCMC) based solver. By defining a likelihood function based on S-parameter equations and incorporating physical priors, we derive optimized impedance values that enhance bandwidth beyond what is reported in the literature. Our flexible pipeline allows efficient tuning of the coupler design. The results demonstrate fractional bandwidths that reach 1.0 for a 9-section coupler, substantially outperforming previous designs. Statistical analysis and convergence tests confirm the robustness of our approach.
\end{abstract}


\section{Introduction}\label{sec:intro}
3 dB Quadrature Hybrid Couplers are essential components that evenly divide the incoming power between the through and coupled output ports. Circuit theory-based design techniques are found in literature, e.g., \cite{chiu2010investigation}, \cite{levy1968synthesis, fathelbab2008synthesis}, \cite{sinha2024solutions}, \cite{liu2022multi}, \cite{shukor2016enhanced}. In particular \cite{levy1968synthesis} has pre-calculated network impedance data to determine impedances of hybrid coupler networks up to nine sections, but the passband performance of the designs are not satisfactory, owing to issues like lower fractional bandwidths, higher passband ripples. Wideband couplers with minimal ripples in the passband are required for astronomical instruments, and a common approach to achieve greater bandwidth is to increase the number of coupler sections. As the number of section increases, the number of dimensions in the non-linear equations grow, and parametric degeneracies occur. Therefore standard solvers are computationally inefficient and in literature most QHDCs are 5-sections or less. We introduce a new approach for designing 3dB (symmetric) QHDCs, addressing this issue. %However, we need to solve for impedances of these couplers with the help of some solver that can solve for impedances fast enough, even for a large number of sections. 

We construct a Markov Chain Monte Carlo (MCMC \cite{foreman2013emcee}) based analysis platform; an optimization program that maximizes a likelihood function with physical priors to derive the most likely impedances that satisfy QHDC design metrics. %Target value constraints on the amplitude of the S-parameters are set, which the likelihood function uses to constrain model equations based on parameterized impedances.

The analysis pipeline is described with a 3-section 3 dB QHDC, before presenting results with larger sections with wideband response. The port assignments in this model, see Fig.~\ref{fig:cascade_twosection}, are: 1- input, 4- isolation, 2- through and 3- coupled respectively. Ports 2 and 3 have a mutual $90^{\circ}$ phase difference and power input from ports 1 or 4 is evenly split across these. Increasing the number of sections makes the passband wider, and we study this trend in detail, accounting for passband ripples. 

\section{Analysis of coupler via unit cell concept}\label{sec:unitcell_analysis}
 The S-parameters for a QHDC must satisfy the conditions specified in Eqn.~\ref{eq:sparams_boundvalues}. For exploring other power splitting ratios beyond 3-dB, these complex values can be adjusted as desired.

\begin{equation}
    S_{11}=0;\;\;S_{21}=\frac{-j}{\sqrt{2}};\;\;S_{31}=\frac{-1}{\sqrt{2}};\;\;S_{41}=0;
    \label{eq:sparams_boundvalues}
\end{equation}

\begin{figure}[h]
    \centering
    \includegraphics[width=1.0\linewidth]{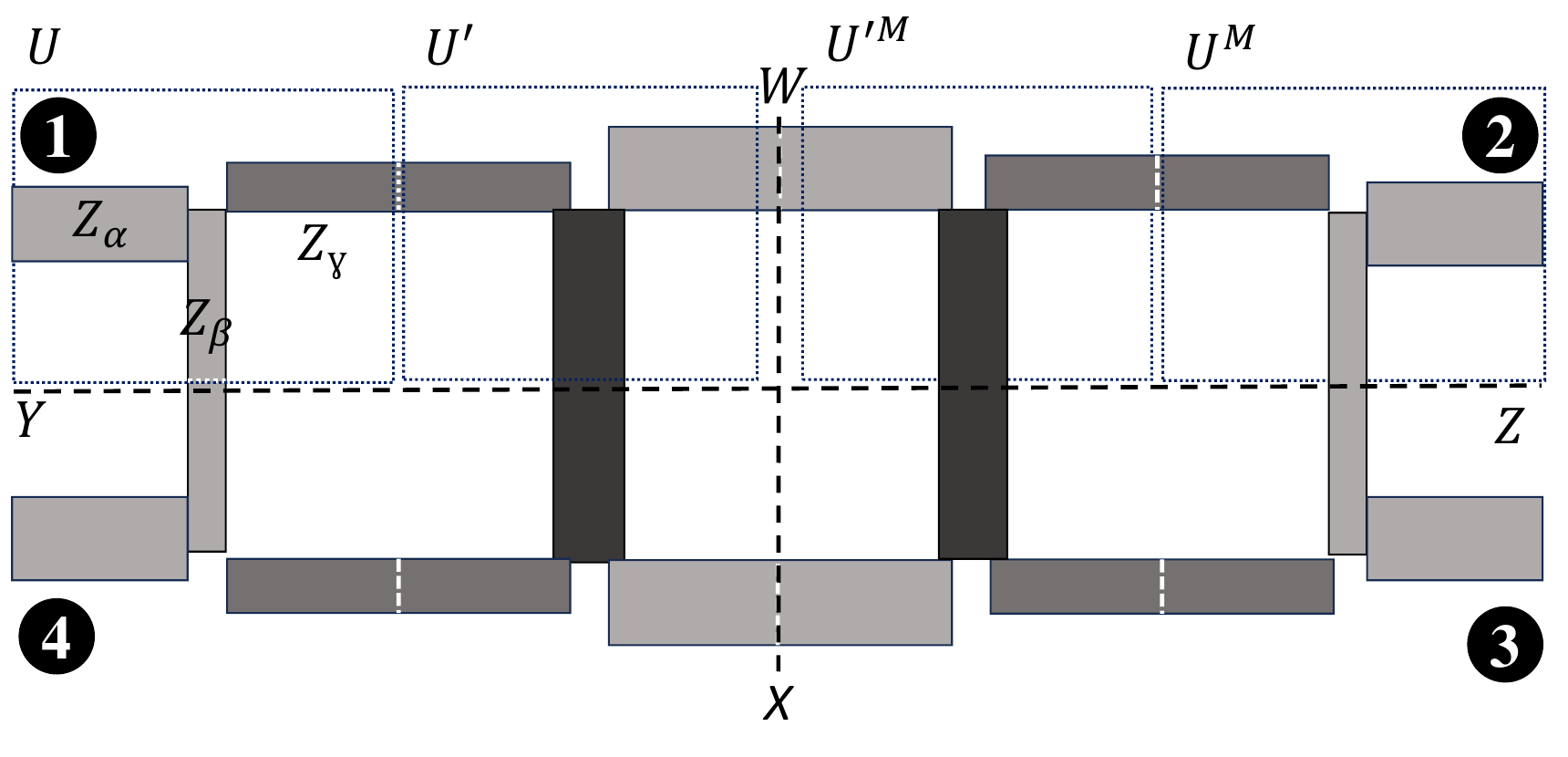}
    \caption{A 3-section QHDC where each arm is $\lambda$/4 long. The ports are labeled as defined in Sec.~\ref{sec:intro}. Cascading of unit-cells to form a 3-section coupler is demonstrated. For the first unitcell from left, impedances being \( Z_\alpha, Z_\beta, Z_\gamma \), in order from left to right. The thickness of the transmission lines give an idea of  its impedance distribution. WX is the line of symmetry for mirror operation shown in Sec.~\ref{sec:evenoddanalysis}. YZ is the line of symmetry for the ABCD operation.}
    \label{fig:cascade_twosection}
\end{figure}

We decompose the total QHDC network into its smallest repeating components, i.e., unit cells. A general analysis pipeline performs Even-Odd analysis on the network of these unit-cells.  ABCD parameters thus obtained can be cascaded to realize a larger \textit{n}-section QHDC ($n = 3$ here, however the process is most general). The unit cell network's ABCD matrix is called $U$, which is the summed even and odd modes' ABCD parameters. This is a function of the impedances of the three arms (\( Z_\alpha, Z_\beta, Z_\gamma \)), the relative permittivity (\( \epsilon_r \)) and the design frequency (\( f \)), see. Fig.~\ref{fig:cascade_twosection}.

% \begin{equation}
%     U = 
%     \begin{bmatrix}
%         A & \quad B \\
%         C & \quad D
%     \end{bmatrix}_{\text{O}}
%     + 
%     \begin{bmatrix}
%         A & \quad B \\
%         C & \quad D
%     \end{bmatrix}_{\text{E}}
%     \label{eq:U_def}
% \end{equation}

\subsection{Even-Odd Analysis of the Unit Cell}\label{sec:evenoddanalysis}
The ABCD parameters for the odd and even modes are calculated for each unit-cell (the unit-cells themselves are distributed in a series/parallel configuration). The total ABCD matrix is derived by cascading these elements. Combining the ABCD parameters from the even and odd modes\footnote{Refer to Appendix~\ref{appendixA} for the equations for 
 even and odd mode unit-cell ABCD. And equation of $A_e$ and $A_o$ for a two section QHDC as demonstration.}, we derive reflection and transmission coefficients, \( \Gamma_e \), \( \Gamma_o \), \( T_e \), and \( T_o \) respectively. These are ultimately used to calculate the QHDC's \( S_{ij} \) given all the parameters defining the unit-cells.
 
For the 3-section QHDC, Fig.~\ref{fig:cascade_twosection}, the coupler of interest is symmetric. That is, the unit-cells at the extremes are identical but mirror images of each other, a factor that has been incorporated into the analysis pipeline and helps in reducing solve times. For the overall ABCD parameters of the network, we multiply the ABCD parameters of the four unit-cells, which are: $U$, $U'$ (both are different owing to their impedances) and their respective mirror images: $U^M$, $U'^M$. See Eqn.~\ref{eq:U} below.
\begin{align}
    U_{\text{final}} &= U(Z_{\alpha}, Z_{\beta}, Z_{\gamma}, \epsilon_r) \times \nonumber \\
    &\quad U'(Z'_{\alpha}, Z'_{\beta}, Z'_{\gamma}, \epsilon_r) \times \nonumber \\
    &\quad U'^M(Z_{\alpha}, Z_{\beta}, Z_{\gamma
 }, \epsilon_r) \times \nonumber \\
    &\quad U^M(Z_{\alpha}, Z_{\beta}, Z_{\gamma}, \epsilon_r)
    \label{eq:U}
\end{align}

%\subsection{Validation of the Analysis pipeline}
We pass impedance values reported by \cite{chiu2010investigation} for a 3-section QHDC into our model and recover the S-parameters which matches with the reference. This serves as validation of the unit-cell cascade process.

\begin{figure}[h!]
    \centering
    \includegraphics[width=1\linewidth]{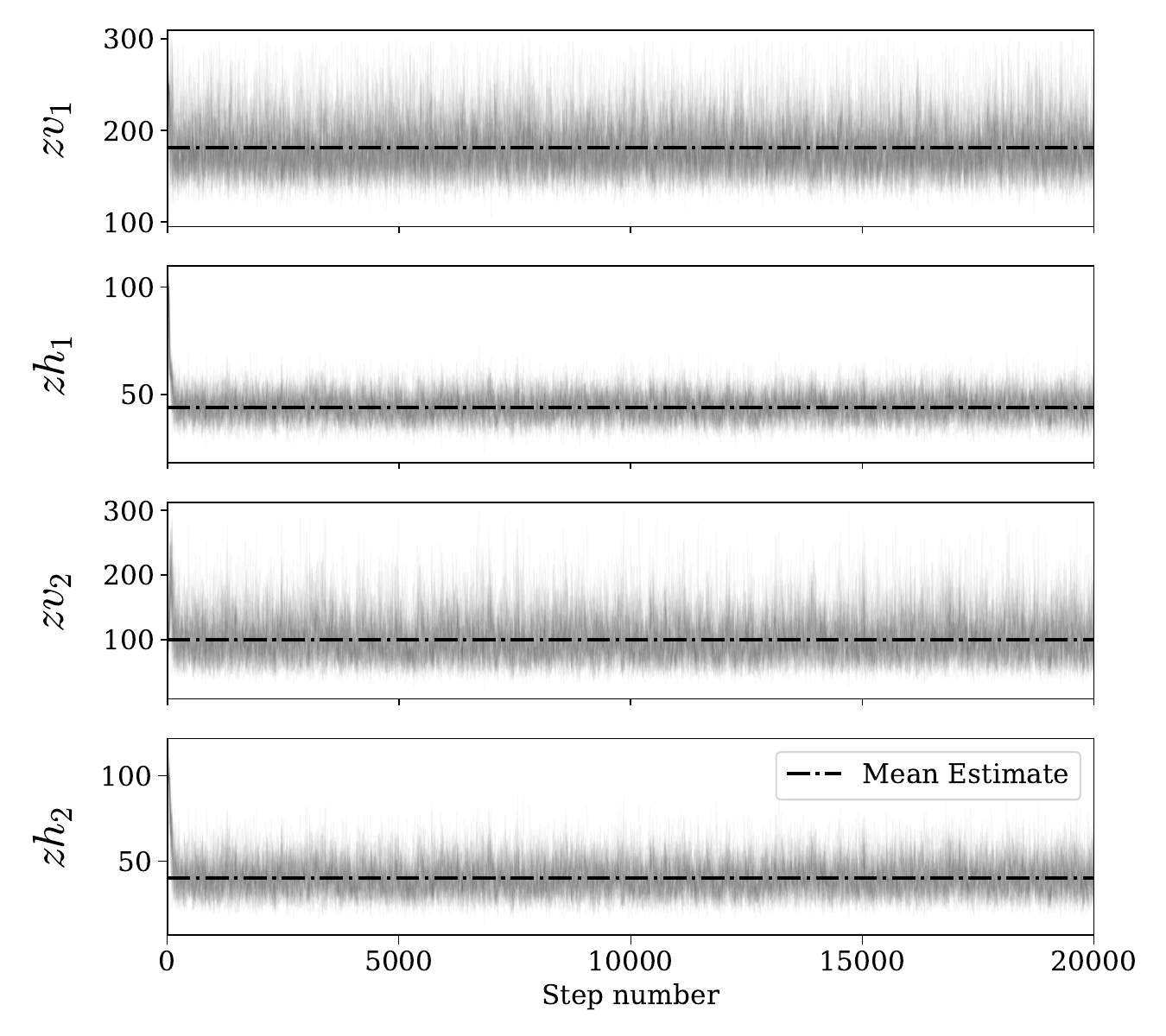}
    \caption{The parameter chain samples are shown, with mean values indicated by a horizontal line. The chains stabilize quickly, starting from a generic guess value.}
    \label{fig:time_series_3section_params}
\end{figure}

\section{Markov Chain Monte Carlo}

In higher-dimensional problems, estimating inferences becomes challenging due to parametric degeneracies and large number of coupled equations. Local extrema can also make it difficult for non-linear solvers which follow deterministic algorithms based on gradient evaluations. Markov Chain Monte Carlo (MCMC) methods are of immense help here, and are workhorses in particle physics and astronomical data analyses. MCMCs include well-known classes of algorithms that draw samples from the probability distributions of parameters being optimized. Optimization is done with a log-likelihood function which is used to evaluate the goodness-of-fit of these samples, given a model. Numerous implementations of MCMC exist and our pipeline\footnote{Currently we are working on upgrades to MACDA to incorporate arbitrary couplers as well, following which a release plan will be designed in a follow-up paper.} utilizes \texttt{emcee}\footnote{\url{https://emcee.readthedocs.io/en/stable/}}; an MIT-licensed, pure-Python implementation of the Goodman-Weare Affine Invariant MCMC Ensemble sampler \cite{goodman2010ensemble}. 

The ``optimize'' function from \texttt{scipy} was tried and it failed to converge for 5-9 section QHDCs for multi frequency points, a requirement for maximizing the bandwidth. The \texttt{scipy}-based optimization functions struggle with convergence given several coupled complex non-linear equations as function of impedances. Number of impedances increase as we move up in sections and progressively the number of parameters become more than the number of equations. Also our posterior distribution of impedances are often multi-modal and standard solvers get stuck in local minima and do not explore the global space for the best solution.

\begin{figure}[!]
    \centering
    \includegraphics[width=1\linewidth]{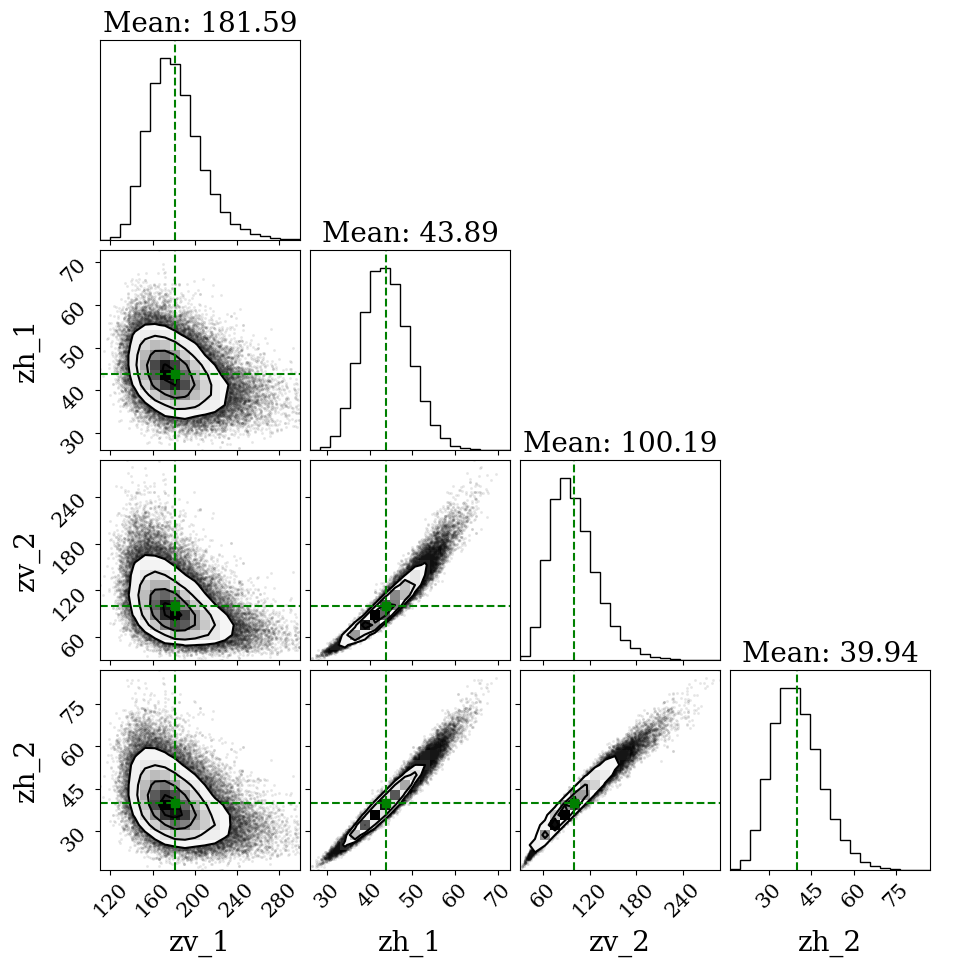}
    \caption{Corner plot of the impedance parameters, showing a skewed posterior. The green line is the mean value. With markers as square for first run outputs from MACDA and circle as values for impedances from \cite{chiu2010investigation}.}
    \label{fig:corner_3section}
\end{figure}

\section{MCMC Assisted Coupler Designer and Analyzer (MACDA) Pipeline}

MACDA\footnote{For the demonstration please email any one of the authors.} is a \texttt{emcee} (python) implementation of MCMC to analyze and solve system of S matrix equations to obtain the most credible impedances for a \textit{n}-section QHDC. It leverages techniques such as parallel computing, code optimization, and equation caching to significantly reduce the computational complexity of solving large-scale S-equations, achieving practical solve times.

\subsection{Demonstration of pipeline with a 3 section QHDC}\label{sec:bandwidth}

% The bandwidth is defined by the frequency points after which the absolute deviation of both \(|S_{21}| \) and \( |S_{31}| \) from -3 dB exceeds 1.8 dB. This defines the passband.
The passband of the QHDC is defined by all contiguous frequencies around the band-center ($\Bar{f}$) for which both \(20 \text{log}_\text{10}(S_{21})\) dB and \(20 \text{log}_\text{10}(S_{31})\) dB are within $\pm$1.8 dB of their nominal -3 dB values.
The left and right edges of the passband therefore define $f_{min}$ and $f_{max}$, or cutoff frequencies.
%Hence, cut-off frequencies are the frequency points on extrema of the passband. 
There are two cut-off frequencies, one  each from \(S_{21}\) and \(S_{31}\) on both side of band-center, and we conservatively select the smallest passband that satisfies the low-ripple threshold. Finally, bandwidth $\text{BW} = f_{max}-f_{min}$, and fractional bandwidth $= \text{BW}/\bar{f}$.
%each for the lower and upper cut-offs on either side of band-center (

\begin{figure}
    \centering
    \includegraphics[width=1\linewidth]{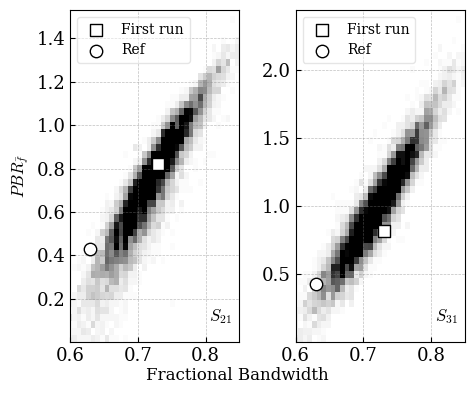}
    \caption{Heatmap of PBR versus fractional bandwidth derived from $S_{21}$ and $S_{31}$. The S-matrices were generated with the impedances drawn from the posterior samples of the MACDA run for a 3-section QHDC. Darker regions indicate more points. `Ref' are the values from \cite{chiu2010investigation}. `First run' corresponds to the impedances from a first MACDA run, without any tuning or optimization. This graph shows that the first run gives sensible and practical numbers. Additionally, it demonstrates the ability of MACDA to explore the phase space of parameters and thus shows the possible landscape of outcomes that can be attained with further need-driven tuning.}
    \label{fig:pbr_kde}
\end{figure}

% Also, \(|S_{21}| + 3 \) and \( |S_{31}| + 3\), at the design frequency, are called the Passband Ripple (PBR) for \(S_{21} \) and \( S_{31} \).
%We also define a single-valued design metric Pass-Band Ripple at $\Bar{f}$ ($\text{PBR}_{\Bar{f}}$) for \(S_{21}\) and \(S_{31}\). 
Functions, \(20 \text{log}_\text{10}(|S_{21}|)\) + 3 dB and \(20 \text{log}_\text{10}(|S_{31}|)\) + 3 dB, over the entire passband is called as Passband Ripple (PBR) of \(S_{21}\) and \(S_{31}\), and are functions of frequency. PBR starts close to the target value at $\bar{f}$ and then diverges away at higher/ lower frequencies. Thus PBR is a function of frequency and at the band-center it is called $\text{PBR}_{\Bar{f}}$. A good QHDC should have low PBR, and in general we expect it to increase as we alter the QHDC design to have larger bandwidths, see Fig.~\ref{fig:pbr21vsbw}. We balance bandwidth expansion against an amount of ripples introduced into the $\text{PBR}_{\Bar{f}}$. This is formally explored in Fig.~\ref{fig:pbr_kde}.

%They serve as a key metric quantifying both the ripple in the passband, as well as the passband itself from when it maximally diverges at the lowest / highest frequencies.
% \textcolor{red}{insert PBR equation. And make it clear here how you can use the whole array, or one number from this (as is needed for Fig 4)}

A 3-section QHDC has four impedance parameters, \( zh_1, zv_1, zh_2, \) and \( zv_2 \), to be solved for at a band-center, e.g., 25 GHz. We constraint the problem at 18 GHz, 25 GHz, 32 GHz to ensure the wideband outcome that is desired.  In these studies we set the dielectric constant as \(\epsilon_r = 4 \). A MACDA user can choose their preferred values for all these parameters (and use more or less number of frequency points). These numbers serve as an example here. The goal is to maximize bandwidth while managing the $\text{PBR}_{\Bar{f}}$. For an acceptable QHDC design: in the passband,  \( 20 \text{log}_\text{10}|S_{11}| \), \( 20 \text{log}_\text{10}|S_{41}| \) $<$ -20 dB and $\text{PBR}_{\Bar{f}}$ for \(S_{21} \) and \( S_{31} \) should be $<$ 1.8 dB.

% The MACDA pipeline takes in various inputs, such as priors on the impedance parameters in log space, which applies uniform prior and returns 1, when the impedances lie between 0 and 300 $\omega$, and 0 elsewhere.

MACDA solves for impedances using a likelihood method, and we use uniform priors. For any impedance $z$, the prior probability is $P(z) = 1$ if $ 300 \Omega > z>0 \Omega$, else $P(z) = 0$. An additional energy conservation prior ensures that the sum of all the four S-parameters (in linear space) squared, remains close to 1.0 with 0.1\% tolerance. These constraints help guide the MCMC process while allowing for a broad parameter search, suitable for this initial exploratory pipeline run.
% not needed, as you explained this in text.
% \begin{equation}
% |\sum_{i=1}^{4} \left| S_{i1} \right|^2 - 1| < 0.001
% \end{equation}

Model functions (derived from analytical cascading of unit-cells) generates the S-parameters as functions of frequency and impedance parameters, providing values for \( S_{11}, S_{21}, S_{31}, \) and \( S_{41} \). To speed up computation of these functions, we use Numba with JIT. Impact of this is summarized in table. \ref{tab:numba_comparison}. With increasingly complex circuits, computation times will increase. For the 3-section case (and up to 9 section), we solve in a reasonable time frame, but 11-section and higher runs may increase in solve times beyond an hour; we will improve these algorithms in future work. The impact of multiprocessing while running MACDA is also shown in table. \ref{tab:core_comparison}.

\renewcommand{\arraystretch}{1.2} % Adjust row height
\begin{table}[h!]
    \caption{Execution Time Comparison Before and After Using Numba for the analyzer function, which brought down time complexities.}
    \label{tab:numba_comparison}
    \centering
    \setlength{\tabcolsep}{10pt} % Adjust column spacing
    \begin{tabular}{|p{1cm}<{\centering}|p{2cm}<{\centering}|p{2cm}<{\centering}|}
        \hline
        \textbf{Sections} & \textbf{Before Numba} & \textbf{After Numba} \\ \hline
        3 & 2.2 sec & 140 $\mu$s \\ \hline
        5 & 8.7 sec & 221 $\mu$s \\ \hline
        7 & 41.2 sec & 302 $\mu$s \\ \hline
        9 & 2.9 mins & 690 $\mu$s \\ \hline
    \end{tabular}
\end{table}

\renewcommand{\arraystretch}{1.2} % Adjust row height
\begin{table}[!h]
    \caption{MACDA Solver Time Comparison: Single Core vs. Multi Core. Given similar inputs to reach a proper solution.}
    \label{tab:core_comparison}
    \centering
    \setlength{\tabcolsep}{8pt} % Adjust column spacing
    \begin{tabular}{|p{1cm}<{\centering}|p{2cm}<{\centering}|p{2cm}<{\centering}|}
        \hline
        \textbf{Sections} & \textbf{Single Core} & \textbf{Multi Core} \\ \hline
        3 & 7 mins & 0.5 mins \\ \hline
        5 & 35 mins & 5 mins \\ \hline
        7 & 4 hrs & 20 mins \\ \hline
        9 & 8.6 hrs & 45 mins \\ \hline
    \end{tabular}
\end{table}

With the model functions, we can define the likelihood ($\mathcal{L}$) as a measure of goodness-of-fit for the proposals generated by the sampler, with target values. We use log of a gaussian likelihood function, as given below in Eqn.~\ref{eq:like_function}. Where \(y(f)\) is the target value from Eqn.~\ref{eq:sparams_boundvalues}, $x_i(f)$ is the S parameter array generated with the sampled impedance, evaluated at frequency $f$. The normalization-scale of the residual  ($y(f) - x_i(f)$) is $y^{err}_i$, which we can reduce, to penalize the model more and vice versa. This can be a function of frequency, though here we use a constant value. In Eqn.~\ref{eq:like_function}, F is the set of frequency solve points, e.g. \(F = \{18, 25, 32\}\) GHz, and  \(S = \{S_{11}, S_{21}, S_{31}, S_{41}\}\).

\begin{align}    
    \mathcal{L} = \sum_{i \in S}\left[- \sum_{f \in F} \frac{1}{2}  \left( \frac{y(f) - x_i(f)}{y_{i, f}^{err}} \right)^2  - \ln \left( \frac{(2 \pi)^{-0.5}}{y_{i, f}^{err}} \right) \right]
    \label{eq:like_function}
\end{align}

Target values for \( 20 \text{log}_\text{10}|S_{11}| \) dB and \( 20 \text{log}_\text{10}|S_{41}|\) dB are set to -60 dB (which can be tuned). The other target values are those from Eqn.~\ref{eq:sparams_boundvalues}. Finally, MCMC input parameters include initial guesses (250, 100, 100, 100 $\Omega$), the number of walkers (30), number of steps (20000), and perturbation (0.1 \%) settings to initialize the Markov chains. Thinning (set to 5, to reduce the correlation between subsequent samples) and a burn-in period is applied to reduce autocorrelation and focus on stable samples. Estimation of burn in period was done with the help of autocorrelation estimators, which gave us uncorrelated samples after 8631 steps. Hence we set our burn-in as 9000 steps. With these settings, we run the MCMC sampling efficiently, leveraging parallel computation to bring down solve time to minutes. An example of the outputs of these walkers upon solution are shown in Fig.~\ref{fig:time_series_3section_params}.

\begin{figure}[!]
    \centering
    \includegraphics[width=1\linewidth]{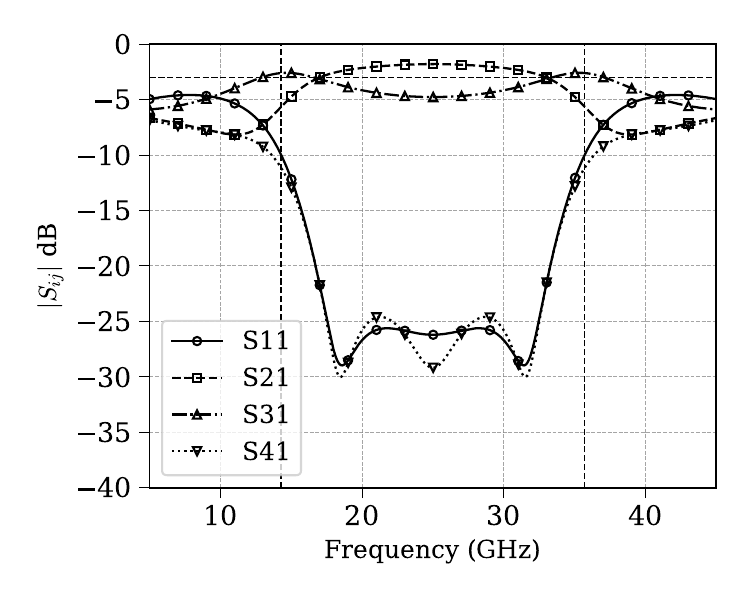}
    \caption{S parameter magnitude versus frequency for our mean values obtained for 3 section run from our MACDA solver. -3dB and 10 GHz line plotted to interpret deviations from design constraints.}
    \label{fig:3section_splot_tuned}
\end{figure}

\begin{figure}[!]
    \centering
    \includegraphics[width=1\linewidth]{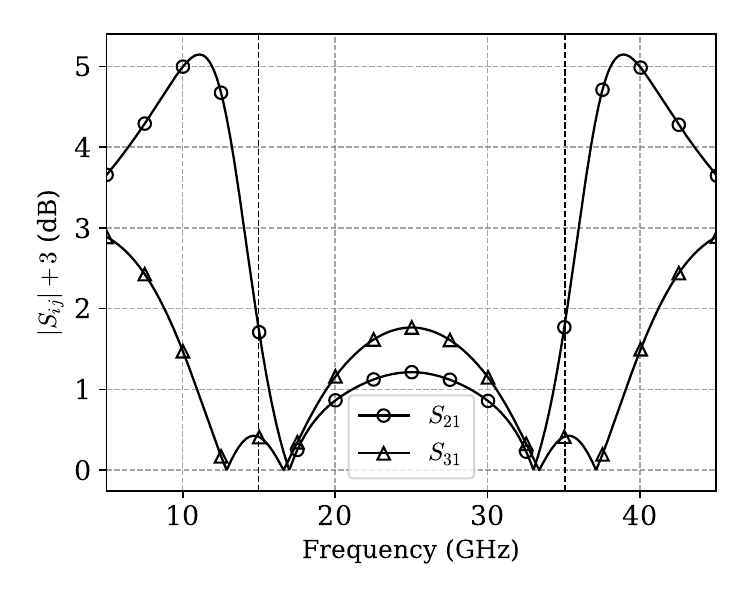}
    \caption{PBR of both $S_{21}$ and $S_{31}$ for 3-section QHDC. The $\text{PBR}_{\Bar{f}}$ for $S_{21}$ is 1.21 dB and for $S_{31}$ is 1.77 dB. With vertical lines at lower and higher cut-off frequency points.}
    \label{fig:3section_s21s31_diff_tuned}
\end{figure}

%In an ideal case (not realizable)  $S_{21}-S_{31} = 0$ over the full passband, vertical dash-dot lines.
\begin{figure}[h!]
    \centering
    \includegraphics[width=1\linewidth]{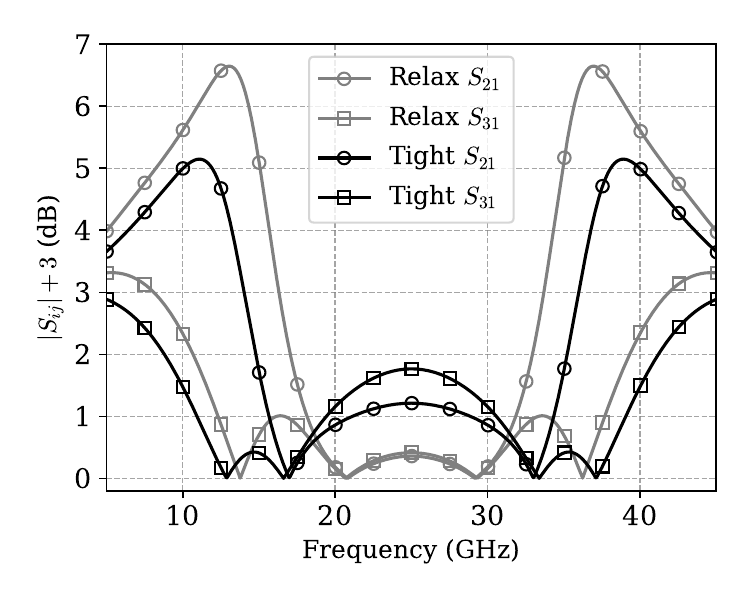}
    \caption{PBR of both $S_{21}$ and $S_{31}$ for 3-section QHDC showing effects of constraint levels, black represents the tight constraint run and light grey represents the relaxed constraint run. The tight constraints yield $\text{PBR}_{\Bar{f}}$ of 0.36 in $S_{21}$ and 0.41 in $S_{31}$, while relaxed constraints yield $\text{PBR}_{\Bar{f}}$ of 0.82 in $S_{21}$ and 1.05 in $S_{31}$. Relaxed constraints help gains in bandwidth but PBR increase.}
    \label{fig:relax_tight_cons}
\end{figure}

\begin{figure}[!]
    \centering
    \includegraphics[width=1\linewidth]{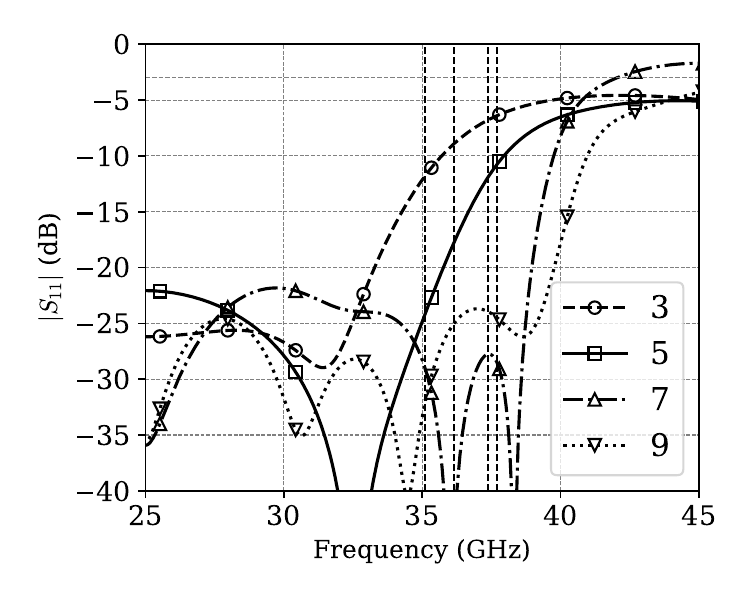}
    \caption{$|S_{11}|$ parameter magnitude for 3, 5, 7, 9 Section QHDC, analyzed with the impedance from  MACDA shows clear expansion in bandwidth with sections. We show the S-plots in the right half ($\geq$ 25 GHz) of the passband to zoom in, and as the plot is symmetric. The vertical dashed lines being the higher cut-off frequency for 3, 5, 7, 9 -section QHDC as we move from left to right. Cut-off frequencies determined based on the criteria outlined in Sec.~\ref{sec:bandwidth}}
    \label{fig:s_11_5_7_9}
\end{figure}

\subsection{Result/Analysis of 3 section run}\label{sec:3section_analysis_}
Convergence test results which include $\hat{R}$ (statistical measure to indicate convergence) $\approx 1$ (for all parameters), clearly indicating convergence of the chains. Respective ``ess'' values (effective sample size) were optimized to a maximum with consecutive runs, to ensure robust statistical sampling. The parameter chain samples are in Fig.~\ref{fig:time_series_3section_params}, and the bi-variate posterior distributions (marginalized over other parameters) are shown in Fig.~\ref{fig:corner_3section}. These demonstrate convergence and the ultimate tolerance-driven solution space explored by MACDA.

We optionally enable the user to further tune the metrics, based on the MACDA outputs. To do this, a GUI based tuner is used to change the impedance values, computing and displaying the new $|S_{ij}(f)|$ plots in real time. From the posteriors, followed by tuning with an aim to enhance bandwidth further, we obtain the impedances as 225.76, 42.17, 94.66, 39.01 $\Omega$ from initial values of 181.5, 43.94, 100.53, 40.06 $\Omega$.  

Directly from MACDA outputs, we sample the posterior distributions of all the impedances to model a family of S-matrices. From these we derive $\text{PBR}_{\Bar{f}}$ and fractional bandwidths. Fig.~\ref{fig:pbr_kde}, shows this exploration of the sample space, and gives the trend (correlations) of design parameters that can be achieved in general. The position of first run values and the reference values are shown for comparison (circle and square markers). %The entire density plot shows the range of PBR and fractional bandwidth the user can play in.

% \renewcommand{\arraystretch}{1.2} % Adjust row height
% \begin{table}[!t]
%     \caption{Summary of Statistical Tests: $\hat{R}$ Statistic, $ess_{tail}$, and $ess_{bulk}$.}
%     \label{tab:3section_statistics}
%     \centering
%     \setlength{\tabcolsep}{6pt} % Adjust column spacing
%     \begin{tabular}{|p{1.2cm}|p{1.2cm}|p{1cm}|p{1.2cm}|p{1.2cm}|}
%         \hline
%         \textbf{Imp} & \textbf{Mean} & \textbf{$\hat{R}$} & \textbf{ess\_bulk} & \textbf{ess\_tail} \\ \hline
%         zv\_1 & 183.525 & 1.01 & 4223.0 & 763.0 \\ \hline
%         zh\_1 & 39.804  & 1.01 & 3675.0 & 896.0 \\ \hline
%         zv\_2 & 76.072  & 1.01 & 3709.0 & 885.0 \\ \hline
%         zh\_2 & 33.391  & 1.01 & 3768.0 & 870.0 \\ \hline
%     \end{tabular}
% \end{table}

% Table 2
\renewcommand{\arraystretch}{1.2} % Adjust row height
\begin{table}[!t]
    \caption{Tuned network Impedance Values for 3-, 5-, 7-, and 9-Section Coupler Runs with the MACDA Pipeline, }
    \label{tab:s_11_5_7_9}
    \centering
    \setlength{\tabcolsep}{8pt} % Adjust column spacing
    \begin{tabular}{|p{1cm}|p{6cm}|}
        \hline
        \textbf{Sections} & \textbf{Impedances ($\Omega$)} \\ \hline
        Three & 225.76, 42.17, 94.66, 39.01 \\ \hline
        Five  & 482.9, 46.4, 200.54, 46.04, 153.78, 43.93 \\ \hline
        Seven & 628, 40.36, 333, 30.58, 136.82, 22.83, 81.15, 19.28\\ \hline
        Nine & 667, 46.11, 645.2, 42.84, 304.85, 39.96, 258.32, 37.57, 199.97, 37.21 \\ \hline
    \end{tabular}
\end{table}

We take the tuned values and study the S-matrix. Fig.~\ref{fig:3section_splot_tuned}, shows $20 \text{log}_\text{10}(|S_{11}|)$ dB and $20 \text{log}_\text{10}(|S_{41}|) \text{ dB} \approx$ -25 dB, and Fig.~\ref{fig:3579_phase_s21s31}A shows a consistent \(90^\circ\) phase offset between the ports 2 and 3, in the entire passband, as desired by Eqn.~\ref{eq:sparams_boundvalues}. Initial values for the impedances had a bandwidth (considering the bandwidth definitions in Sec.~\ref{sec:bandwidth}) of 18.34 GHz (fractional bandwidth of 0.73). With its passband from 15.87 GHz to 34.21 GHz. As we tuned this QHDC design to enhance the current bandwidth, we observed an increase in PBR. Fig.~\ref{fig:3section_s21s31_diff_tuned}, shows the final PBR for both $S_{21}$ and $S_{31}$ (below our 1.8 dB threshold), and the final passband is thus obtained, see Tab.~\ref{tab:comp_bw_table}.

%{Add a couple of sentences here on how you use the S21-S31 plots to define acceptable ripple and passband. This information is present earlier, but it's important to repeat this as one is looking at all these final graphs.} From our earlier bandwidth definition, where we take the 

\subsection{Effect of relaxed and tight constraints}

The $y_{err}$ residual normalization parameter can ``tighten'' the MACDA runs. A larger $y_{err}$ enables looser constraints and broader exploration of phase-space than a smaller value. Using $y_{err}=10^{-6}$, results in a tighter exploration of the parameter space and yields the ripple offset plot shown in Fig. \ref{fig:relax_tight_cons}. This is an extra design-knob, which can also be made frequency dependent to fine tune the impedances so as to better attain a user specific S-parameters.

\subsection{Design for 5, 7, 9 -section QHDC}\label{sec:5section_analysis}

Following similar strategies in the previous section, we have also designed a 5, 7 and 9-section QHDC at 25 GHz band-center. The final impedance values, see Tab.~\ref{tab:s_11_5_7_9}, are obtained by adjusting all the input parameters to the MACDA run and fine tuning its parameters to meet our requirements.

\begin{figure}[!]
    \centering
    \includegraphics[width=1\linewidth]{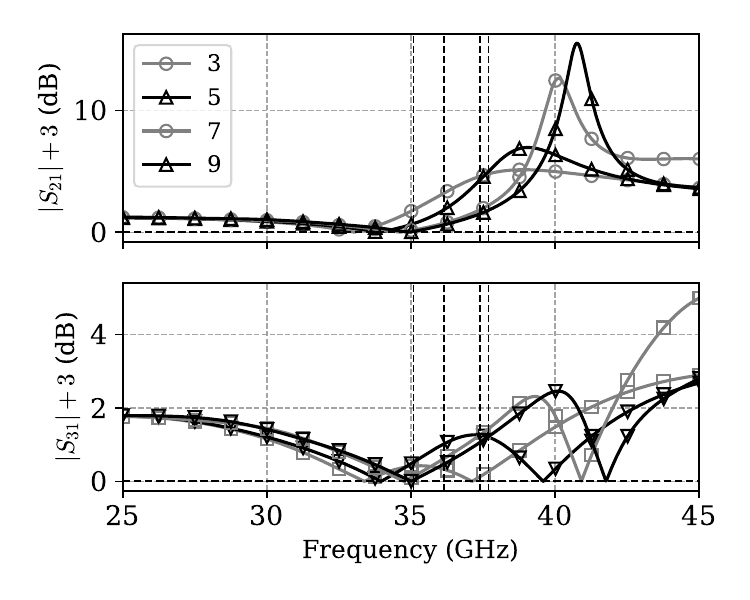}
    \caption{PBR in $S_{21}$ (top) and $S_{31}$ (bottom) for 3, 5, 7, 9 -section QHDC with impedances generated from MACDA. The vertical dashed lines being the higher cut-off frequency for 3, 5, 7, 9 -section QHDC as we move from left to right. We show the S-plots in the right half ($\geq$ 25 GHz) of the passband to zoom in, and as the plot is symmetric.}
    \label{fig:s_21_5_7_9}
\end{figure}

\begin{figure}
    \centering
    \includegraphics[width=1\linewidth]{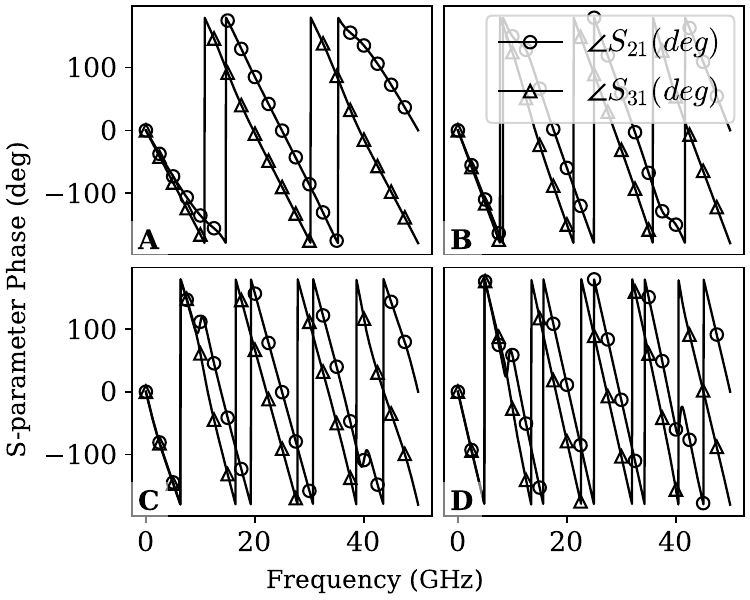}
    \caption{S parameter phase versus frequency for (A) 3-section QHDC with phase being 0.42\textdegree\ for $S_{21}$ and -89.59\textdegree\ for $S_{31}$, (B) 5-section QHDC with phase being -179.41\textdegree\ for $S_{21}$ and 90.58\textdegree\ for $S_{31}$, (C) 7-section QHDC with phase being 0.77\textdegree\ for $S_{21}$ and -89.23\textdegree\ for $S_{31}$, (D) 9-section QHDC with phase being -179.06\textdegree\ for $S_{21}$ and 90.94\textdegree\ for $S_{31}$, all at the band-center; with impedances generated from MACDA.}
    \label{fig:3579_phase_s21s31}
\end{figure}

In Fig.~\ref{fig:s_11_5_7_9}, we show $|S_{11}|$ variation for 3, 5, 7 and 9 -section QHDC, and we see that the passband increases significantly as we move to higher sections. From Fig.~\ref{fig:s_21_5_7_9}, the bandwidths are, 0.80, 0.89, 0.99, 1.01 for a 3, 5, 7 and 9 -section QHDC respectively. The phase of $S_{21}$ and $S_{31}$ (in degrees), for all four QHDCs are presented in Fig.~\ref{fig:3579_phase_s21s31}. We observe a consistent \(90^\circ\) phase offset between the ports 2 and 3, in the entire passband.

The 5-section QHDC shows an improvement in bandwidth from 3-section QHDC, leading to a fractional bandwidth of 0.89. With the observed $\text{PBR}_{\Bar{f}}$ under 1.8 dB respectively for $S_{21}$ and $S_{31}$, see Tab.~\ref{tab:comp_bw_table}. We have presented an in-depth comparison of MACDA generated designs with the reference text values in Sec.\ref{sec:comp_with_lit}. The 7-section QHDC further improves the bandwidth, compared to the 5-section QHDC, leading to a fractional bandwidth of 0.99, with $\text{PBR}_{\Bar{f}}$ for $S_{21}$ and $S_{31}$ also staying under 1.8 dB.

\renewcommand{\arraystretch}{1.2} % Adjust row height
\begin{table}[!t]
    \caption{Comparison of $BW^{[2]}$ from \cite{chiu2010investigation} and $BW^{*}$ which are the values from this Work alongwith their corresponding $C_{i1}^{[*]}$, which is the $\text{PBR}_{\Bar{f}}$ for $S_{21}$ and $S_{31}$ respectively. $f1_{*}$ \& $f2_{*}$ refers to the cut off frequencies for Bandwidth. Dashes implies new results presented in this work.}
    \label{tab:comp_bw_table}
    \centering
    \setlength{\tabcolsep}{8pt} % Adjust column spacing
    \begin{tabular}{|>{\raggedright\arraybackslash}p{0.01cm}|>{\centering\arraybackslash}p{0.6cm}|>{\centering\arraybackslash}p{0.5cm}|>{\centering\arraybackslash}p{1.4cm}|>{\centering\arraybackslash}p{0.4cm}|>{\centering\arraybackslash}p{0.4cm}|>{\centering\arraybackslash}p{0.4cm}|>{\centering\arraybackslash}p{0.4cm}|}
        \hline
        \textbf{} & \textbf{$BW^{[2]}$} & \textbf{$BW^{*}$}  & \textbf{$f1_{*}$ \& $f2_{*}$} & \textbf{$C_{21}^{[2]}$} & \textbf{$C_{21}^{[*]}$} & \textbf{$C_{31}^{[2]}$} & \textbf{$C_{31}^{[*]}$} \\ \hline
        3 & 0.63 & 0.80 & 14.98 - 35.09 & 0.43 & 1.21 & 0.50 & 1.77 \\ \hline
        5 & 0.797 & 0.89 & 13.90 - 36.12 & 1.01 & 1.19 & 1.35 & 1.79 \\ \hline
        7 & -- & 0.99 & 12.66 - 37.39 &  - & 1.24 &  - & 1.77\\ \hline
        9 & -- & 1.01 & 12.33 - 37.69 & - & 1.25 &  - & 1.79 \\ \hline
    \end{tabular}
\end{table}

However for a 9 -section QHDC design, in order to keep the $\text{PBR}_{\Bar{f}}$ for $S_{21}$ and $S_{31}$ under 1.8 dB, we could not increase the fractional bandwidth beyond 1.01, see Tab.~\ref{tab:comp_bw_table} for more details. A more relaxed QHDC design in terms of further enhanced bandwidth by adding more of the PBR and relaxing the target constraints, is presented in Sec.\ref{sec:relax_9section}. The excess PBR that is introduced into the design, should be theoretically reduced by implementing a model with a chebyshev coupler response. It might not be possible to attain reduction in PBR by implementing the current Butterworth type of coupler response. However Chebyshev designs will have lesser bandwidth compared to a Butterworth design. Which brings in the aspect of tuning. This stays a topic of interest for future work.

\begin{figure}[!]
    \centering
    \includegraphics[width=1\linewidth]{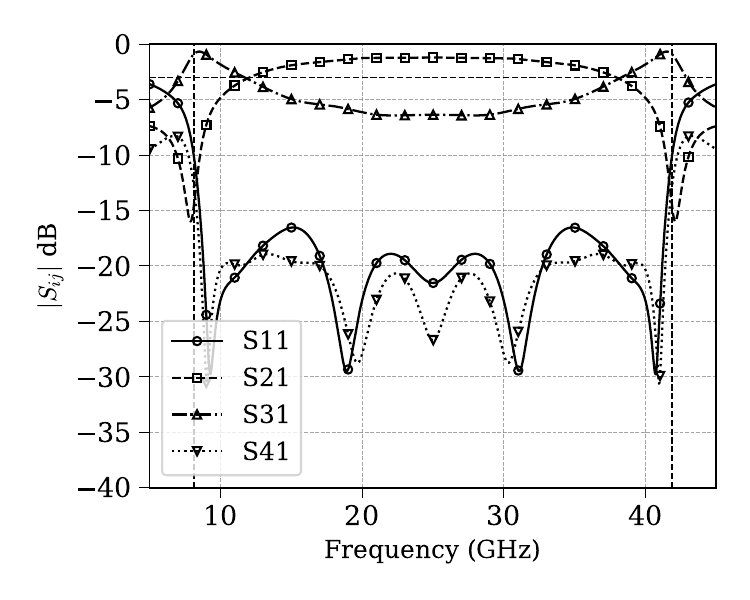}
    \caption{S-parameter magnitude versus frequency for a relaxed 9 section QHDC. -3dB and 10 GHz line plotted to interpret deviations from design constraints.}
    \label{fig:9section_s11}
\end{figure}
\begin{figure}[!]
    \centering
    \includegraphics[width=1\linewidth]{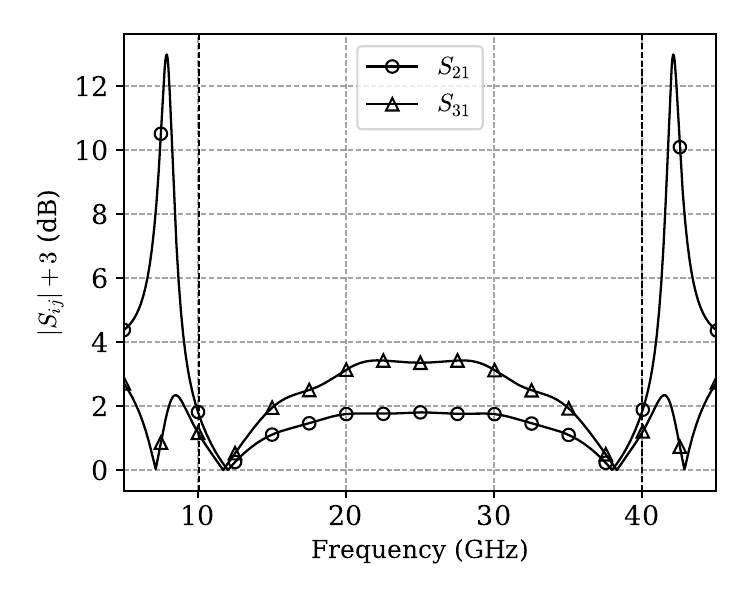}
    \caption{PBR of $S_{21}$ and $S_{31}$ for 9-section QHDC. The PBR for $S_{21}$ is 1.80 dB and for $S_{31}$ is 3.35 dB at the band-center. Vertical lines show lower and higher cut-off frequency points.}
    \label{fig:9section_s21s31}
\end{figure}

\section{Comparison with literature}\label{sec:comp_with_lit}

We perform a qualitative analysis of our 3 and 5 -section QHDC designs by comparing our bandwidth and $\text{PBR}_{\Bar{f}}$ to those of the network impedances derived in \cite{chiu2010investigation}. The reference paper reports impedance values for a 3-section coupler as 211.7, 42.7, 74.8, 36.65 $\Omega$, and for a five-section coupler as 407.1, 49.05, 206.2, 48.37, 165.1, 48.05 $\Omega$. The discussion is summarized in Table \ref{tab:comp_bw_table}, with overall improvements in bandwidth at the expense of increasing the PBR and hence the $\text{PBR}_{\Bar{f}}$, within specific tolerance.

%\textcolor{red}{Just quoting their Z's vs yours is not sufficient. Add a line or two on how much did your ripple and passband change from theirs. Basically point to the table you have later, and add a nice sentence to summarize the 3-section comparison here.}

% \begin{figure}
%     \centering
%     \includegraphics[width=1\linewidth]{photos/bw_variation.pdf}
%     \caption{Bandwidth variation for 3, 5, 7 and 9 -section QHDC. With fractional bandwidth on the left y-axis and passband ripple variation on the right y-axis, with respective $S_{21}$ or $S_{21}$ labels refer to their corresponding passband ripples. Where [2] refers to reference text and * is the design from this work.}
%     \label{fig:bw_variation}
% \end{figure}

\begin{figure}
    \centering
    \includegraphics[width=1\linewidth]{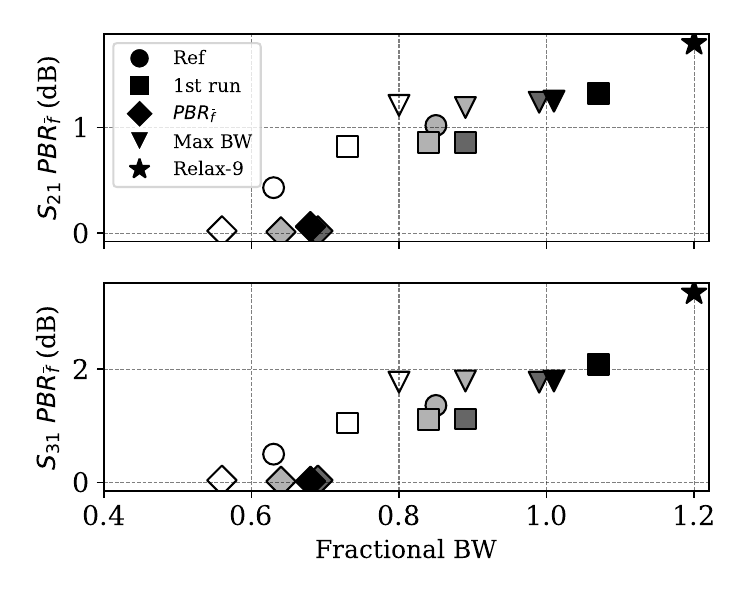}
    \caption{Plot of $\text{PBR}_{\Bar{f}}$ for \(S_{21}\) (top) and \(S_{31}\) (bottom) versus fractional bandwidth for QHDCs with varying numbers of sections (marker symbols). Namely, 3-sections: white, 5-sections: light-gray, 7-sections: dark-gray, 9-sections: black symbols respectively. `Ref' refers to values from \cite{chiu2010investigation}, and all others are from MACDA runs. `Relax-9' is the 9 -section QHDC that was made in Sec.~\ref{sec:relax_9section}, for tunability demonstration, if target constraints can be relaxed.}
    \label{fig:pbr21vsbw}
\end{figure}

\subsection{3-section QHDC analysis}In Table \ref{tab:comp_bw_table}, we observe that our design for the 3-section QHDC demonstrates improved in-band performance compared to the results in \cite{chiu2010investigation}, in terms of fractional bandwidth as 0.80 (over 0.63 in \cite{chiu2010investigation}). However, this improvement comes at the expense of higher $\text{PBR}_{\Bar{f}}$ as 1.21 and 1.77 for $S_{21}$ and $S_{31}$ (over 0.43 and 0.50 respectively, for \cite{chiu2010investigation}).

\subsection{5-section QHDC analysis}
For the 5-section QHDC, our bandwidth improves to 0.89 from 0.797 in \cite{chiu2010investigation}. Similar to the 3 -section QHDC, higher $\text{PBR}_{\Bar{f}}$ are introduced, which are 1.19 and 1.79 for $S_{21}$ and $S_{31}$ (over 1.01 and 1.35 respectively, for \cite{chiu2010investigation}). 

\vspace{0.6em}
There is no 7-section QHDC reported in the reference paper \cite{chiu2010investigation}, to compare with. And neither is there for 9-section QHDC.

\section{Designing a relaxed 9 -section QHDC}\label{sec:relax_9section}
Obtaining high fractional bandwidth, like 1.2, is not possible with 3, 5, 7 -section coupler, as could be seen in Fig.~\ref{fig:pbr21vsbw}. But it is possible for a 9 -section QHDC with a relaxed constraint. By changing our requirements for an acceptable QHDC design from what is defined in Sec.~\ref{sec:bandwidth}, we will attempt to obtain a design with a fractional bandwidth of 1.2. Followed by analyzing the rise in $\text{PBR}_{\Bar{f}}$ of $S_{21}$ and $S_{31}$. We will also allow the $20 \text{log}_\text{10}(|S_{11}|)$ dB and $20 \text{log}_\text{10}(|S_{41}|)$ dB to rise up to -15 dB, if required. With the tuner, we find the new network impedances as 667, 44.49, 715.2, 40.95, 337.42, 39.42, 328.32, 32.17, 223.21, 29.91 $\Omega$. We show the absolute S-parameter plot in Fig.~\ref{fig:9section_s11}, where $20 \text{log}_\text{10}(|S_{11}|)$ dB and $20 \text{log}_\text{10}(|S_{41}|)$ dB $\approx$ -20 dB. From Fig.~\ref{fig:9section_s21s31}, we obtain the fractional bandwidth as 1.20 with cut-off frequencies at 10.06 to 40 GHz (bandwidth of 29.93 GHz). The $\text{PBR}_{\Bar{f}}$ for $S_{21}$ was quantified as 1.80 dB and for $S_{31}$ it increased to 3.35 dB.

The designs in literature are often accompanied with large ripple offsets, which we have reduced successfully to under 2 dB for up to nine section in Sec.~\ref{sec:3section_analysis_} and Sec.~\ref{sec:5section_analysis}. This section aimed to show the possibility to gain in bandwidth if specific user scenario allows more relaxed target requirements.

\section{Demonstration of Tunability with MACDA}

We present various runs' data in Appendix.~\ref{sec:data_run_tuning} to demonstrate tunability. Also Fig.~\ref{fig:pbr21vsbw} shows the variation of $\text{PBR}_{\Bar{f}}$ for $S_{21}$ and $S_{31}$ against the fractional bandwidth. It is clear from the plots that the solver combined with the tuner aids the user to tune the QHDC, in either reducing the $\text{PBR}_{\Bar{f}}$, which drags down the bandwidth as well. Or we can attempt to raise the bandwidth but it increases the $\text{PBR}_{\Bar{f}}$ in both $S_{21}$ and $S_{31}$. With the relaxed 9-section QHDC design, we finally find a value which gives us a fractional bandwidth of 1.2.

%The dashed line represents the assumed ripple limit, while the vertical line marks a fractional bandwidth of 1.

%We have tuned and reduced the value of the ripple for the nine section from 5.1 dB initially, to 3.42 dB, without compromising $S_{11}$ or $S_{41}$. These are new findings and further reduction is unlikely without $S_{11}$ increasing beyond -20 dB. In summary, the presented nine section coupler design gives us a large bandwidth of 1.07 from 11.69 GHz to 38.35 GHz.

\section{Future work}
We have developed solutions for a \(n\)-section symmetric couplers, where n is odd, and the coupler provides a half-power split. Moving forward, we plan to extend MACDA to accommodate even-numbered sections, asymmetric configurations, and custom power splits. We aim to reduce the solve times to improve computational efficiency for higher sections. The pipeline currently enforces a Butterworth-type response in the passband. However to reduce -3 dB ripple offset in higher section QHDCs ($>$9), we shall implement models to design Chebyshev type couplers. Possibility to design different types of devices, is also considered. Thus eventually MACDA will enable a vast range of devices to be designed, based on desired frequency dependent complex S-parameters that the user inputs.

\section*{Acknowledgment}
The authors are grateful for the helpful discussions with Dr. Saurabh Singh, Raman Research Institute and Sonia Ghosh, University of Groningen.

%{\appendices
%\section*{Proof of the First Zonklar Equation}
%Appendix one text goes here.
% You can choose not to have a title for an appendix if you want by leaving the argument blank
%\section*{Proof of the Second Zonklar Equation}
%Appendix two text goes here.}

% \newpage

% \section{Biography Section}
% If you have an EPS/PDF photo (graphicx package needed), extra braces are
%  needed around the contents of the optional argument to biography to prevent
%  the LaTeX parser from getting confused when it sees the complicated
%  $\backslash${\tt{includegraphics}} command within an optional argument. (You can create
%  your own custom macro containing the $\backslash${\tt{includegraphics}} command to make things
%  simpler here.)
 
% \vspace{11pt}

% \bf{If you include a photo:}\vspace{-33pt}
% \begin{IEEEbiography}[{\includegraphics[width=1in,height=1.25in,clip,keepaspectratio]{fig1}}]{Michael Shell}
% Use $\backslash${\tt{begin\{IEEEbiography\}}} and then for the 1st argument use $\backslash${\tt{includegraphics}} to declare and link the author photo.
% Use the author name as the 3rd argument followed by the biography text.
% \end{IEEEbiography}

% \vspace{11pt}

% \bf{If you will not include a photo:}\vspace{-33pt}
% \begin{IEEEbiographynophoto}{John Doe}
% Use $\backslash${\tt{begin\{IEEEbiographynophoto\}}} and the author name as the argument followed by the biography text.
% \end{IEEEbiographynophoto}

% \vfill

\newpage
\onecolumn
\appendices
\section{ABCD Matrices of Unit-Cell and equations for $A_e$ and $A_o$ for 2-section QHDC for demonstration.}\label{appendixA}

\vspace{1.2em}
To provide additional clarity, the derived even and odd ABCD matrix for the unit-cell are given below. 

\vspace{1.2em}

\[
\begin{bmatrix}
    A & \quad B \\
    C & \quad D
\end{bmatrix}_{\text{O}} =
\begin{bmatrix}
    \frac{-Z_{\alpha} \cdot S^2}{Z_{\gamma}} + \left( \frac{Z_{\alpha} \cdot S}{Z_{\beta} \cdot T} + C \right) \cdot C
    & \quad
    I \cdot Z_{\alpha} \cdot S \cdot C + I \cdot Z_{\gamma} \cdot \left( \frac{Z_{\alpha} \cdot S}{Z_{\beta} \cdot T} + C \right) \cdot S
    \\
    \left( \frac{-I \cdot C}{Z_{\beta} \cdot T} + \frac{I \cdot S}{Z_{\alpha}} \right) \cdot C + \frac{I \cdot S \cdot C}{Z_{\gamma}}
    & \quad
    I \cdot Z_{\gamma} \cdot \left( \frac{-I \cdot C}{Z_{\beta} \cdot T} + \frac{I \cdot S}{Z_{\alpha}} \right) \cdot S + C^2
\end{bmatrix}
\]

\[
\begin{bmatrix}
    A & \quad B \\
    C & \quad D
\end{bmatrix}_E =
\begin{bmatrix}
    \frac{-Z_{\alpha} \cdot S^2}{Z_{\gamma}} + \left( \frac{-Z_{\alpha} \cdot S \cdot T}{Z_{\beta}} + C \right) \cdot C
    & \quad
    I \cdot Z_{\alpha} \cdot S \cdot C + I \cdot Z_{\gamma} \cdot \left( \frac{-Z_{\alpha} \cdot S \cdot T}{Z_{\beta}} + C \right) \cdot S
    \\
    \left( \frac{I \cdot C \cdot T}{Z_{\beta}} + \frac{I \cdot S}{Z_{\alpha}} \right) \cdot C + \frac{I \cdot S \cdot C}{Z_{\gamma}}
    & \quad
    I \cdot Z_{\gamma} \cdot \left( \frac{I \cdot C \cdot T}{Z_{\beta}} + \frac{I \cdot S}{Z_{\alpha}} \right) \cdot S + C^2 
\end{bmatrix}
\]

\begin{equation}
    \text{and,} \; S = \sin(A), \; C = \cos(A), \; T = \tan(A)\notag
\end{equation}

\begin{equation}
    \text{where,} \; A = 3.14159 \times 10^{-11} \cdot f \notag 
\end{equation} \notag

\vspace{1.2em}

The equation for $A_e$ for a 2-section  
QHDC is given below with same substitutions as the above equation. 3-section QHDC equations, which was discussed in detail in Sec.~\ref{sec:bandwidth}, was not presented due to the sheer size of the equations.

\begin{align}
    &\Bigg[ \Big( \frac{I \cdot C \cdot T}{z_{v_2}} + \frac{I \cdot S}{z_{h_1}} \Big) C + \frac{I \cdot S \cdot C}{z_{h_1}} \Bigg] \Bigg[ I \cdot z_0 \cdot S \cdot C + I \cdot z_{h_1} \Big( \frac{-z_0 \cdot S \cdot T}{z_{v_1}} + C \Big) S \Bigg] \nonumber \\
    &+ \Bigg[ \frac{-z_{h_1} \cdot S \cdot T}{z_{v_2}} + C \Bigg] C - S^2 \Bigg[ \frac{-z_0 \cdot S^2}{z_{h_1}} + \Big( \frac{-z_0 \cdot S \cdot T}{z_{v_1}} + C \Big) C \Bigg] \nonumber \\
    &\Bigg[ \Big( \frac{-z_{h_1} \cdot S \cdot T}{z_{v_1}} + C \Big) C - \frac{z_{h_1} \cdot S^2}{z_0} \Bigg] + \Bigg[ \frac{-z_0 \cdot S^2}{z_{h_1}} + \Big( \frac{-z_0 \cdot S \cdot T}{z_{v_1}} + C \Big) C \Bigg] \nonumber \\
    &\Bigg[ I \cdot z_{h_1} \Big( \frac{-z_{h_1} \cdot S \cdot T}{z_{v_2}} + C \Big) S + I \cdot z_{h_1} S \cdot C \Bigg] + \Bigg[ I \cdot z_0 \cdot S \cdot C + I \cdot z_{h_1} \Big( \frac{-z_0 \cdot S \cdot T}{z_{v_1}} + C \Big) S \Bigg] \nonumber \\
    &\cdot \Bigg[ I \Big( \frac{I \cdot C \cdot T}{z_{v_2}} + \frac{I \cdot S}{z_{h_1}} \Big) S + C^2 \Bigg].
\end{align}

\vspace{1.2em}

The equation for $A_o$ for a 2-section QHDC is given below with same substitutions as the above equation. 

\begin{equation}
\begin{aligned}
& \Bigg[ \Bigg( \frac{-I \cdot \cos(A)}{z_{v_2} \cdot T} + \frac{I \cdot \sin(A)}{z_{h_1}} \Bigg) \cdot C + \frac{I \cdot \sin(A)}{z_{h_1}} \cdot C \Bigg] \cdot \Bigg[ I \cdot z_0 \cdot \sin(A) \cdot C + I \cdot z_{h_1} \Bigg( \frac{z_0 \cdot \sin(A)}{z_{v_1} \cdot T} + C \Bigg) \cdot \sin(A) \Bigg] \\
& + \Bigg[ \frac{-z_{h_1} \cdot \sin(A)}{z_{v_2} \cdot T} + C \Bigg] \cdot C - \sin(A)^2 \cdot \Bigg[ \frac{-z_0 \cdot \sin(A)^2}{z_{h_1}} + \Bigg( \frac{-z_0 \cdot \sin(A)}{z_{v_1} \cdot T} + C \Bigg) \cdot C \Bigg] \\
& \cdot \Bigg[ \Bigg( \frac{-z_{h_1} \cdot \sin(A)}{z_{v_1} \cdot T} + C \Bigg) \cdot C - \frac{z_{h_1} \cdot \sin(A)^2}{z_0} \Bigg] + \Bigg[ \frac{-z_0 \cdot \sin(A)^2}{z_{h_1}} + \Bigg( \frac{-z_0 \cdot \sin(A)}{z_{v_1} \cdot T} + C \Bigg) \cdot C \Bigg] \\
& \cdot \Bigg[ I \cdot z_{h_1} \Bigg( \frac{-z_{h_1} \cdot \sin(A)}{z_{v_2} \cdot T} + C \Bigg) \cdot \sin(A) + I \cdot z_{h_1} \cdot \sin(A) \cdot C \Bigg] + \Bigg[ I \cdot z_0 \cdot \sin(A) \cdot C + I \cdot z_{h_1} \Bigg( \frac{-z_0 \cdot \sin(A)}{z_{v_1} \cdot T} + C \Bigg) \cdot \sin(A) \Bigg] \\
& \cdot \Bigg[ I \Bigg( \frac{I \cdot \sin(A) \cdot T}{z_{v_2}} + \frac{I \cdot \sin(A)}{z_{h_1}} \Bigg) \cdot \sin(A) + C^2 \Bigg]. \notag 
\end{aligned}
\end{equation}

\newpage

\section{Run data obtained for tunability Demonstration of MACDA}\label{sec:data_run_tuning}

% 3-section table
\begin{table*}[h]
\caption{3 -Section QHDC design runs}
\label{tab:3_section}
\centering
\begin{tabular}{|l|l|l|l|l|}
\hline
Run & Impedances ($\Omega$) & $S_{21}$,$S_{31}$ ($\text{PBR}_{\Bar{f}}$) (dB) & BW (Frac. BW) & Cut-off frequency (GHz) \\ \hline
From Reference \cite{chiu2010investigation} & 211.7, 42.7, 74.8, 36.65 & 0.43, 0.50 & 15.74 (0.63) & 17.14, 32.88 \\ \hline
First Run & 181.5, 43.94, 100.53, 40.06 & 0.82, 1.05 & 18.34 (0.73) & 15.86, 34.21 \\ \hline
Lowest PBR & 158.9, 37.77, 65.3, 32.93 & 0.02, 0.04 & 14.08 (0.56) & 17.97, 32.05 \\ \hline
Highest BW & 225.76, 42.17, 94.66, 39.01 & 1.21, 1.77 & 20.11 (0.80) & 14.98, 35.09 \\ \hline
\end{tabular}
\end{table*}

% 5-section table
\begin{table*}[h]
\caption{5 -Section QHDC design runs}
\label{tab:5_section}
\centering
\begin{tabular}{|l|l|l|l|l|}
\hline
Run & Impedances ($\Omega$) & $S_{21}$,$S_{31}$ $\text{PBR}_{\Bar{f}}$ (dB) & BW (Frac.) & Cut-off (GHz) \\ \hline
From Reference \cite{chiu2010investigation} & 407.1, 49.05, 206.2, 48.37, 165.1, 48.05 & 1.01, 1.35 & 21.39 (0.86) & 14.34, 35.73 \\ \hline
First Run & 364.68, 48.49, 193.77, 46.04, 154.82, 44.97 & 0.86, 1.11 & 20.89 (0.84) & 14.59, 35.48 \\ \hline
Lowest ripple & 314.7, 51.88, 211.48, 50.73, 129.82, 47.57 & 0.01, 0.02 & 15.92 (0.64) & 17.07, 32.98 \\ \hline
Highest BW & 482.9, 46.4, 200.54, 46.04, 153.78, 43.93 & 1.19, 1.79 & 22.22 (0.89) & 13.90, 36.12 \\ \hline
\end{tabular}
\end{table*}

% 7-section table
\begin{table*}[h]
\caption{7 -Section QHDC design runs}
\label{tab:7_section}
\centering
\begin{tabular}{|l|l|l|l|l|}
\hline
Run & Impedances ($\Omega$) & $S_{21}$,$S_{31}$ $\text{PBR}_{\Bar{f}}$ (dB) & BW (Frac.) & Cut-off (GHz) \\ \hline
First Run & 478, 41.4, 288, 31.1, 141.77, 22.57, 74.9, 19.8 & 0.86, 1.12 & 22.32 (0.89) & 13.86, 36.19 \\ \hline
Lowest ripple & 478, 41.4, 274.2, 30.84, 96.01, 22.31, 69.95, 19.8 & 0.02, 0.04 & 17.32 (0.69) & 16.37, 33.68 \\ \hline
Highest BW & 628, 40.36, 333, 30.58, 136.82, 22.83, 81.15, 19.28 & 1.24, 1.77 & 24.72 (0.99) & 12.66, 37.39 \\ \hline
\end{tabular}
\end{table*}

% 9-section table
\begin{table*}[h]
\caption{9 -Section QHDC design runs}
\label{tab:9_section}
\centering
\begin{tabular}{|l|l|l|l|l|}
\hline
Run & Impedances ($\Omega$) & $S_{21}$,$S_{31}$ $\text{PBR}_{\Bar{f}}$ (dB) & BW (Frac.) & Cut-off (GHz) \\ \hline
First Run & 517, 47.46, 670.2, 42.84, 329.85, 39.96, 283.32, 37.03, 223.21, 37.21 & 1.32, 2.09 & 26.68 (1.07) & 11.69, 38.38 \\ \hline
Lowest ripple & 667, 46.92, 545.2, 43.11, 279.85, 40.23, 258.3, 37.57, 113.7, 37.21 & 0.06, 0.02 & 16.97 (0.68) & 16.55, 33.52 \\ \hline
Highest BW & 667, 46.11, 645.2, 42.84, 304.85, 39.96, 258.32, 37.57, 199.97, 37.21 & 1.25, 1.79 & 25.36 (1.01) & 12.33, 37.69 \\ \hline
Relax case & 667, 44.49, 715.2, 40.95, 337.42, 39.42, 328.32, 32.17, 223.21, 29.91 & 1.80, 3.35 & 29.93 (1.20) & 10.06, 40.00 \\ \hline
\end{tabular}
\end{table*}

\end{document}